\documentclass[10pt]{article}

\usepackage{amsmath}
\usepackage{amssymb}

\usepackage{graphicx}

\usepackage{cite}

\usepackage{color}

\makeatletter
\renewcommand{\@biblabel}[1]{\quad#1.}
\makeatother

\date{}

\begin{document}

\begin{flushleft}
{\Large
\textbf{Bone in vivo: Surface mapping technique}
}
\\
Yifang Fan$^{1,\ast}$,
Yubo Fan$^{2}$,
Zhiyu Li$^{3}$,
Changsheng Lv$^{1}$
\\
\bf{1} Center for Scientific Research, Guangzhou Institute of Physical Education, Guangzhou 510500, P.R. China
\\
\bf{2} Bioengineering Department, Beijing University of Aeronautics and Astronautics, Beijing 100191, P.R. China
\\
\bf{3} College of Foreign Languages, Jinan University, Guangzhou 510632, P.R. China
\\
$\ast$ E-mail: tfyf@gipe.edu.cn
\end{flushleft}

\section*{Abstract}
Abstract: Bone surface mapping technique is proposed on the bases of two kinds of uniqueness of bone in vivo, (i) magnitude of the principal moments of inertia, (ii) the direction cosines of principal axes of inertia relative to inertia reference frame. We choose the principal axes of inertia as the bone coordinate system axes. The geographical marks such as the prime meridian of the bone in vivo are defined and methods such as tomographic reconstruction and boundary development are employed so that the surface of bone in vivo can be mapped. Experimental results show that the surface mapping technique can both reflect the shape and help study the surface changes of bone in vivo. The prospect of such research into the surface shape and changing laws of organ, tissue or cell will be promising.

\section*{Introduction}
The shape of bone is the adaptive result of bone in the mechanical and physiological environment~\cite{Buckwalter,Nomura,Hugate,Canalis}. A map, the visual representation of our real world symbol model, can reveal not only the spatial structure properties of an object but also the changes in time series~\cite{Legendre,Anas,Barnes}. The mapping of bone surface, therefore, is used as an approach to study the adaptability of bone morphology. Mapping technique has become an even more powerful and useful method to do scientific research. Mapping and flattening techniques have also been widely used in medical research~\cite{Wang,Suri,Saenz,Bates,Morley,Cuntz}. They are both concerned with the development methods. The flattening technique develops the three-dimensional object to a two-dimensional one~\cite{Bennis} while the mapping technique plays an essential role in interpreting the surface structure of an object~\cite{Haker}. The advancements and improvements of three-dimensional imaging of bone in vivo~\cite{Keyak,Bouxsein,Matthews} have brought better data collection methods of bone surface, but the mapping techniques have not been systematically explored with satisfactory results.

To map the bone surface, some geographic marks and geographic coordinates such as the bone surface prime meridian, the equator line or the contour should be identified~\cite{Davies}. They should be defined on the bone's coordinate system. In order to study the bone's changes caused by the external factors, it is fundamental to set up a bone coordinate system when mapping the bone surface. In our study, we have set up a coordinate system with uniqueness on the principal axes of inertia of the bone in vivo. The coordinate system is thus employed to determine the prime meridian, and the average radius of the tomographic boundary is employed to determine the contour to map the bone surface. This means that the bone surface mapping technique could present an alternate approach to study the bone's morphology.

\section*{Standardized coordinate system of bone}
The CT image of bone in vivo can generate the principal moments of inertia and the direction cosines~\cite{Coburn}. Suppose the moment of inertia of the bone's tissue relative to its center of mass is a constant; then the magnitude of the bone's principal moments of inertia will be determined by its shape and mass distribution. The inertia tensor suggests that we can always find a group of coordinate systems where three products of inertia will be nil at the same time~\cite{Hinrichs,Nikravesh}. The magnitudes of these three principal moments of inertia could come out with three results: (i) all three are equal; (ii) two out of three are equal and (iii) each one is different from the other. When the object is homogeneous, in the first case, it is a sphere; in the second, an ellipsoid, a cube, a cylinder or a rectangular. When the bone's location and orientation relative to inertia reference frame (i.e. the coordinate system is defined by the CT coordinate system) are fixed, in the first case, there are numerous principal axes of inertia, while in the second case, the orientation of one principal axis can be determined, but not the other two. Therefore, in the first and second case, the principal moments of inertia have nothing to do with their principal axes of inertia. In the third case, however, the direction cosines of the bone's principal axis of inertia relative to inertia reference frame are unique, which means a one-to-one corresponding relation between the direction cosines of the bone's principal axes of inertia and its shape.

If the bone is defined as a collection of elements of volume $\Delta V$, the element's position can be represented by $(x_{oi},y_{oi},z_{oi})$ (where $o$ is located in the center of mass ), then the magnitude of the principal moments of inertia can be represented respectively by:
\begin{equation}
\label{eq1}
\begin{array}{llllll}
I_{x}=\sum\left(y^{2}_{oi}+z^{2}_{oi}\right)\rho_{i}\Delta V\\
I_{y}=\sum\left(x^{2}_{oi}+z^{2}_{oi}\right)\rho_{i}\Delta V\\
I_{z}=\sum\left(x^{2}_{oi}+y^{2}_{oi}\right)\rho_{i}\Delta V,
\end{array}
\end{equation}
where $\rho$ is the density, $\Delta V=\Delta x\Delta y\Delta z$, $\Delta x$ and $\Delta y$ are the pixel sizes and $\Delta z$ the layer distances of CT images.

Let the angular displacements of the bone that take turns to rotate around axes $x,y,z$ be $\alpha,\beta,\gamma$. According to Equations ~\ref{eq1}, we can set up the following equation:
\begin{equation}
\label{eq2}
\begin{array}{llllll}
\left(I_{y}-I_{z}\right)_{\alpha}=\sum[\left(y_{oi}\cos\alpha-z_{oi}\sin\alpha\right)^{2}-\left(y_{oi}\sin\alpha+z_{oi}\cos\alpha\right)^{2}]\rho_{i}\Delta V\\
\left(I_{x}-I_{z}\right)_{\beta}=\sum[\left(x_{oi}\cos\beta+z_{oi}\sin\beta\right)^{2}-\left(x_{oi}\sin\beta-z_{oi}\cos\beta\right)^{2}]\rho_{i}\Delta V\\
\left(I_{x}-I_{y}\right)_{\gamma}=\sum[\left(x_{oi}\cos\gamma-y_{oi}\sin\gamma\right)^{2}-\left(x_{oi}\sin\gamma+y_{oi}\cos\gamma\right)^{2}]\rho_{i}\Delta V.
\end{array}
\end{equation}

Next, differentiate Eq.~\ref{eq2}, and let
\begin{displaymath}
\frac{d\left(I_{y}-I_{z}\right)_{\alpha}}{d\alpha},\hspace{0.50cm}\frac{d\left(I_{x}-I_{z}\right)_{\beta}}{d\beta},\hspace{0.50cm}\frac{d\left(I_{x}-I_{y}\right)_{\gamma}}{d\gamma},
\end{displaymath}
which will generate the following equation:

\begin{equation}
\label{eq3}
\begin{array}{llllll}
\alpha=\frac{1}{2}\arctan\left(\frac{2\sum \left(y_{oi}z_{oi}\rho_{i}\Delta V\right)}{\sum \left(y^{2}_{oi}\rho_{i}\Delta V\right)-\sum \left(z^{2}_{oi}\rho_{i}\Delta V\right)} \right) \hspace{0.50cm} (A)\\
\beta=\frac{1}{2}\arctan\left(\frac{2\sum \left(x_{\alpha i}z_{\alpha i}\rho_{i}\Delta V\right)}{\sum \left(x^{2}_{\alpha i}\rho_{i}\Delta V\right)-\sum \left(z^{2}_{\alpha i}\rho_{i}\Delta V\right)} \right)\hspace{0.50cm} (B)\\
\gamma=\frac{1}{2}\arctan\left(\frac{2\sum \left(x_{\beta i}y_{\beta i}\rho_{i}\Delta V\right)}{\sum \left(x^{2}_{\beta i}\rho_{i}\Delta V\right)-\sum \left(y^{2}_{\beta i}\rho_{i}\Delta V\right)} \right)\hspace{0.50cm} (C).
\end{array}
\end{equation}

It is apparent that when and ONLY when $I_{x}\neq I_{y}\neq I_{z}$ will Eq.~\ref{eq3} have a set of solutions. $\sum y_{i}z_{i}\rho_{i}\Delta V$, $\sum z_{i}z_{i}\rho_{i}\Delta V$ and $\sum x_{i}y_{i}\rho_{i}\Delta V$ are three products of inertia of inertia tensor. Within the range of $[0,\pi]$, according to Eqs.~\ref{eq2} and~\ref{eq3}, the limited rotations can always turn three products of inertia into zero at the same time.

The bone's shape is asymmetrical and its structure is anisotropic~\cite{Biewener,Hans,Kleber,Ketcham,Fink}. Eq.~\ref{eq3} exposes the direction cosines of principal axes of inertia relative to the inertia reference frame of the bone characterization are unique. As a result, the coordinate system set upon the principal moment of inertia can not only depict the position and orientation of the bone, but also verify the bone surface shape and its changes when making a quantitative analysis. Eqs.~\ref{eq2} and~\ref{eq3} also suggest that we can set up a coordinate system whose coordinate origin is located arbitrarily at the center of mass of the bone. After limited rotations, the coordinate axes is positioned on the principal axes of inertia.

\section*{Mapping of the Bone surface}
CT scanning simplifies the bone as a collection of elements of volume $\Delta V$ (different densities). When performing an isotropic scanning (where pixel size must be the same as the layer distance), the volume of $\Delta V$ is a constant. The position of $\Delta V$ relative to the center of mass differs from one another. When the result of Eq.~\ref{eq3} is replaced for that in Eq.~\ref{eq2} accordingly and when the bone's coordinate system is positioned on the principal axes of inertia, the rotation will change the original CT image. It is necessary, then, to reconstruct the new image, which can be performed by the following equation:
\begin{equation}
\label{eq4}
\begin{array}{llllll}
x_{i}=trunc\left(\frac{x_{oi}-\min(x_{oi})}{\Delta x}\right)\Delta x+\min(x_{oi})\\
y_{i}=trunc\left(\frac{y_{oi}-\min(y_{oi})}{\Delta y}\right)\Delta y+\min(y_{oi})\\
z_{i}=trunc\left(\frac{z_{oi}-\min(z_{oi})}{\Delta z}\right)\Delta z+\min(z_{oi}),
\end{array}
\end{equation}
where $(x_{oi},y_{oi},z_{oi})$ stands for the position of $\Delta V$ after rotation, $(x_{i},y_{i},z_{i})$ for that of the reconstructed tomogram and $trunc()$ for a function that truncates a number to an integer by removing the fractional part of the number. To keep the CT images isotropic, $\Delta d_{x}=\Delta d_{y}=\Delta d_{z}$ is defined in Eq.~\ref{eq4}, and its pixal size and layer distance are kept the same of those of the original image. According to Eq.~\ref{eq4}, when keeping the $\Delta V$ to be in a cube, the new CT image after rotation remains to be closed and continuous.

CT scanning divides the bone surface into a collection of the tomographic image boundaries. In this way, the mapping of the bone surface has become an issue to develop the tomographic boundary, making it necessary to detect and draw the tomographic boundary. Before a new CT scanning, the equipment is reset, i.e. the gray value of the air is set as zero. The scanned tomographic images of bone are processed by Eq.~\ref{eq4}, and their boundaries are drawn by the following equation:
\begin{equation}
\label{eq5} \rho(x,y)_{z}=\left\{
\begin{array}{llllll}
\rho(x,y)_{z} \hspace{0.50cm} \rho(x,y)_{z}>0,\rho(x+1,y)_{z}=0\\
\rho(x,y)_{z} \hspace{0.50cm} \rho(x,y)_{z}>0,\rho(x-1,y)_{z}=0\\
\rho(x,y)_{z} \hspace{0.50cm} \rho(x,y)_{z}>0,\rho(x,y+1)_{z}=0\\
\rho(x,y)_{z} \hspace{0.50cm} \rho(x,y)_{z}>0,\rho(x,y-1)_{z}=0\\
0 \hspace{0.50cm} other,
\end{array}\right.
\end{equation}
where $z$ is the number of sequence of tomogram, $(x,y)_{z}$ the position of $\Delta V$ relative to the tomographic center of mass and $\rho(x,y)_{z}$ the density of $(x,y)_{z}$.

When mapping the bone surface, the bone will be "cut" - from a cylindrical surface to a rectangle, or a rhombus. However it is cut, its ultimate area would be the same. But when it is a rectangle, there is only one. How to cut the bone into a rectangle? The two principal axes of inertia (the minimal and maximal principal moment of inertia) form a plane. On this plane, the  bone surface boundary is called the prime meridian, which is used as the surface cutting line to develop the bone surface. The following equation will make it happen:
\begin{equation}
\label{eq6} i=\left\{
\begin{array}{llllll}
0 \hspace{0.50cm} x^{i}_{z}=x^{c}_{z}, y^{i}_{z}>y^{c}_{z}\\
1 \hspace{0.50cm} x^{i}_{z}-x^{c}_{z}=-\Delta x, y^{i}_{z}>y^{c}_{z}\\
2 \hspace{0.50cm} x^{i}_{z}-x^{c}_{z}=-2\Delta x, y^{i}_{z}>y^{c}_{z}\\
\cdots \hspace{0.50cm}\\
n-1 \hspace{0.50cm} x^{i}_{z}-x^{c}_{z}=2\Delta x, y^{i}_{z}>y^{c}_{z}\\
n \hspace{0.50cm} x^{i}_{z}-x^{c}_{z}=\Delta x, y^{i}_{z}>y^{c}_{z},

\end{array}\right.
\end{equation}
where $(x^{i}_{z},y^{i}_{z})$ is the position of the $\Delta V$ at the tomographic boundary relative to the tomographic center of mass, $(x^{c}_{z},y^{c}_{z})$ the position of tomographic center of mass relative to the inertia reference frame and $i$ the sequence of $\Delta V$ of the boundary after being cut.

Eq.~\ref{eq6} sequences the $\Delta V$s at the tomographic boundary which has been cut. The average radium of the tomographic boundary perpendicular to the minimal (maximal) principal moment of inertia is defined as the "sea level", which is used as a datum line so that the tomographic boundary can be developed by the following equation:
\begin{eqnarray}
\label{eq7}
p(x,y)_{z}=p\left(i+x^{c}_{z},h^{i}_{z}\right)_{z},
\end{eqnarray}
where $z$ shares the same definition of that in Eq.~\ref{eq5}, $h^{i}_{z}=r^{i}_{z}-\overline{r}_{z}$, $r^{i}_{z}=\sqrt{(x^{i}_{z}-x^{c}_{z})^{2}+(y^{i}_{z}-y^{c}_{z})^{2}}$ and $\overline{r}_{z}=\frac{\sum r^{j}_{z}}{n}$, $(x^{c}_{z},y^{c}_{z})$ and $(x^{i}_{z},y^{i}_{z})$ have the same definitions as those in Eq.~\ref{eq6}.

Eqs.~\ref{eq4}-~\ref{eq7} have developed the closed surface into an open three-dimensional curved one with the properties of a contour. The three-dimensional curved map of the bone surface can be further developed into a two-dimensional plane. We can, however, translate the bone surface into a plane with the help of the following equation:
\begin{eqnarray}
\label{eq8}
p(x,y)=p\left(\int \left(\sqrt{\l^{2}(i)+Z^{2}_{i}}\right)di,\int \left(\sqrt{l^{2}(j)+Z^{2}_{j}}\right)dj\right),
\end{eqnarray}
where $p(x,y)$ presents the position of $\Delta V$ in a plane whose surface has been flattened, and
\begin{displaymath}
l(i)=\sqrt{(z_{i}-z_{i-1})^{2}+Z^{2}_{i}},\hspace{0.30cm}
Z_{i}=\int |z_{i}-z_{i-1}|di,\hspace{0.30cm}
l(j)=\sqrt{(z_{j}-z_{j-1})^{2}+Z^{2}_{j}},\hspace{0.30cm}
Z_{j}=\int |z_{j}-z_{j-1}|dj,\\
\end{displaymath}
$z$ stands for the value of contour on $(i,j)$ in bone surface mapping.

Eq.~\ref{eq8} suggests that the mapping of the bone surface actually serves as a simulation of the bone surface structure. It is a space model of an image symbol to represent the bone surface. It shares the consistency with the real body of the bone surface structure.

\section*{Experiment}
Prior to our study, the Ethic Committee of Guangzhou Institute of Physical Education has proved our study and the participant has provided fully informed consent to participate in this study by signing a written consent form. From January 2008 to August 2009, we followed the track of Guangdong Provincial Youth Team of Male Wrestlers by using a 64 slice scanner (Brilliance 64, Philips Medical Systems). Excluding team members who left the team halfway and those with injuries, in January 2008 and August 2009, we were able to scan and collect the data of a 25-year-old wrestler's sesamoid bones beneath the head of the first metatarsal bone of both feet. See Table 1 for information of the sesamoid bones.

In Table~\ref{t1}, the volumes of left and right foot's internal and external sesamoid bone have changed by $0.26\%$, $-1.98\%$, $0.38\%$ and $-1.32\%$ respectively; their surface areas have changed by $0.22\%$, $-0.41\%$, $0.54\%$ and $-0.23\%$ respectively and their density by $0.85\%$, $1.45\%$, $2.40\%$ and $3.77\%$ respectively.

\begin{figure}[!ht]
\begin{center}
\begin{tabular}{cccc}
 \includegraphics[width=11.8cm]{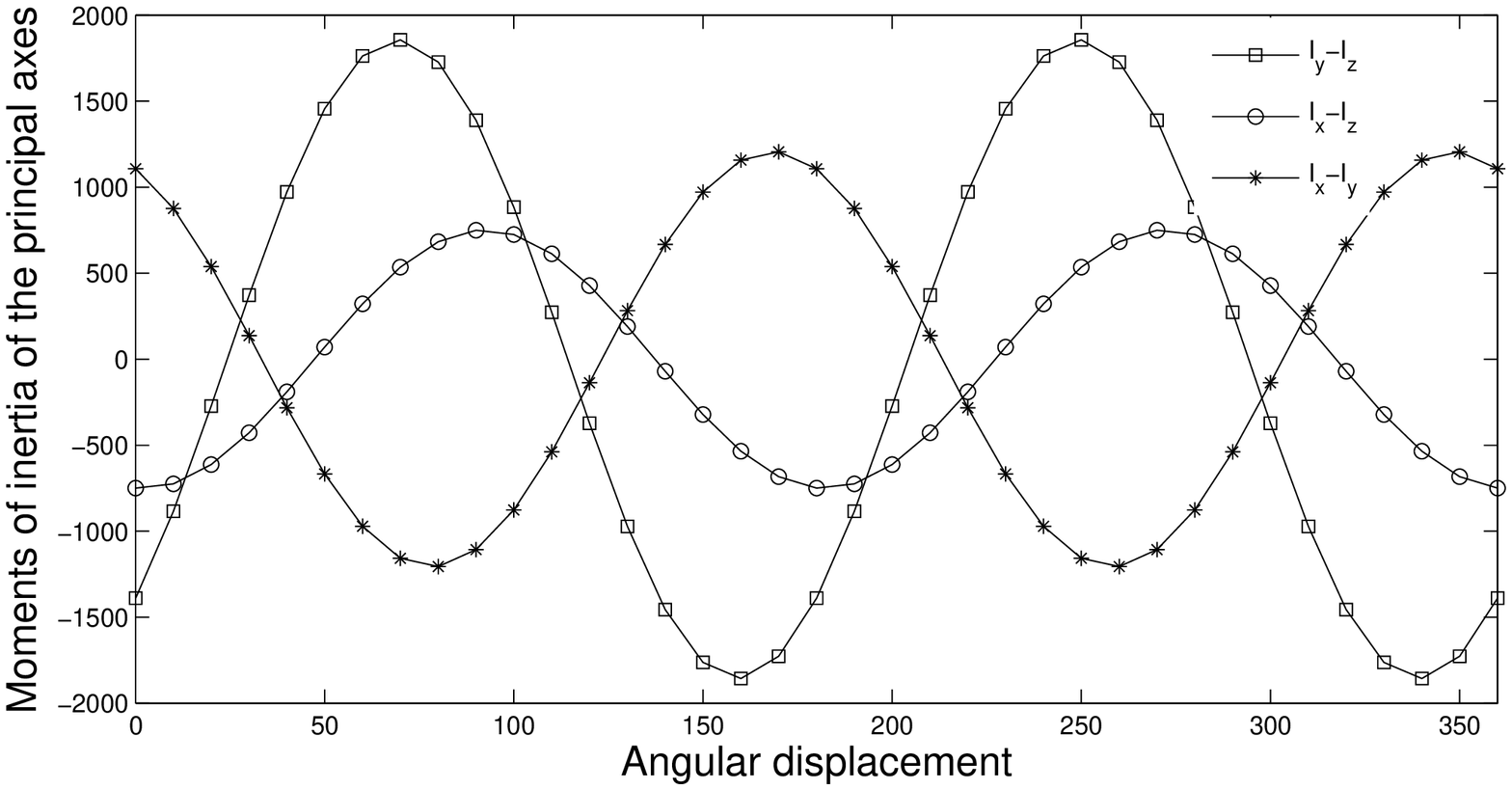}
\end{tabular}
\caption{\label{fig1} The variations of the axis moment of inertia of the first measurement of the wrestler's lower external sesamoid bone of the first metatarsal bone accompanying the changes of right foot.}
\end{center}
\end{figure}

Let's make a mapping analysis to the right foot's external sesamoid beneath the head of the first metatarsal bone. First, rotate the first-time scanned sesamoid bone around axis $x$ from 0 to 180 degrees. According to Eq.~\ref{eq2}, the variations of $(I_{y}-I_{z})_{\alpha}$, $(I_{x}-I_{z})_{\beta}$ and $(I_{x}-I_{y})_{\gamma}$ are shown in Fig.~\ref{fig1}, indicating that within the range of rotation from 0 to 180 degrees, an extremum exists in Eq.~\ref{eq2}. We can get the result of $\alpha=157.73$ when the extremum is calculated by Eq.~\ref{eq3}A. After the sesamoid bone rotates around axis $x$ at $157.63$ degree, it then rotates around axis $y$. The calculation by Eq.~\ref{eq3}B generates the result of $\beta=8.92$. When the sesamoid bone rotates around axis $y$ at $8.92$ degree, the calculation of Eq.~\ref{eq3}C generates the result of $\gamma=172.07$. When the coordinate system of the sesamoid bone rotate around axes $xyz$ at $157.63/8.92/172.07$ degree respectively, the axes of of the sesamoid bone coincide with those of the principal moments of inertia.

It's shown that the morphologically asymmetric and heterogeneously distributed bone has the uniqueness of direction cosines of their principal moments of inertia. An arbitrary coordinate system set upon the bone's center of mass can make every coordinate axis coincide with the principal moment of inertia by coordinate transformation, i.e. the proposed principal axis's coordinate system has its uniqueness. This method can be applied to both the homogeneous asymmetric geometry and the heterogeneous asymmetric geometry.

\begin{figure}[!ht]
\begin{center}
\begin{tabular}{cccc}
 \includegraphics[width=6.8cm]{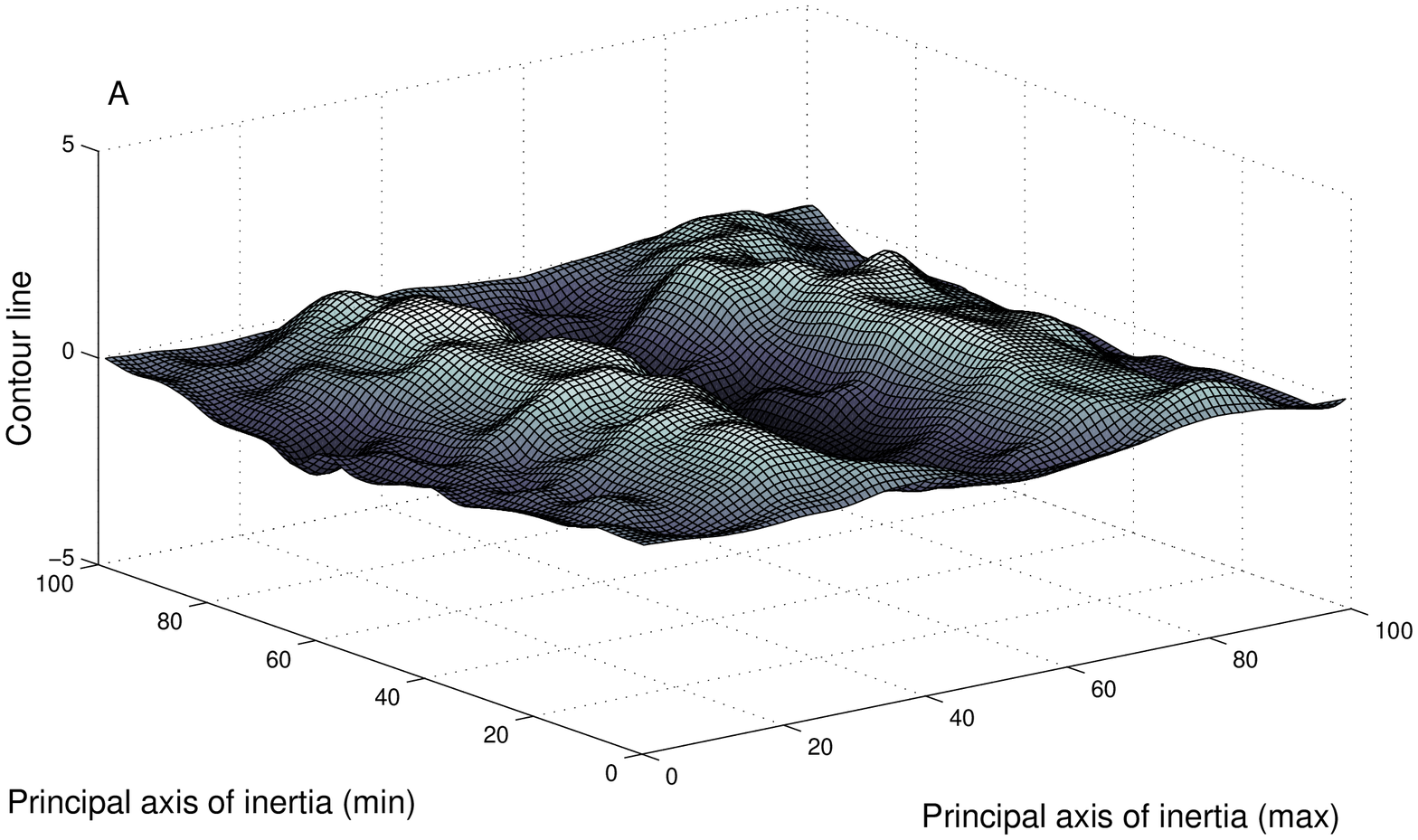}&
 \includegraphics[width=6.8cm]{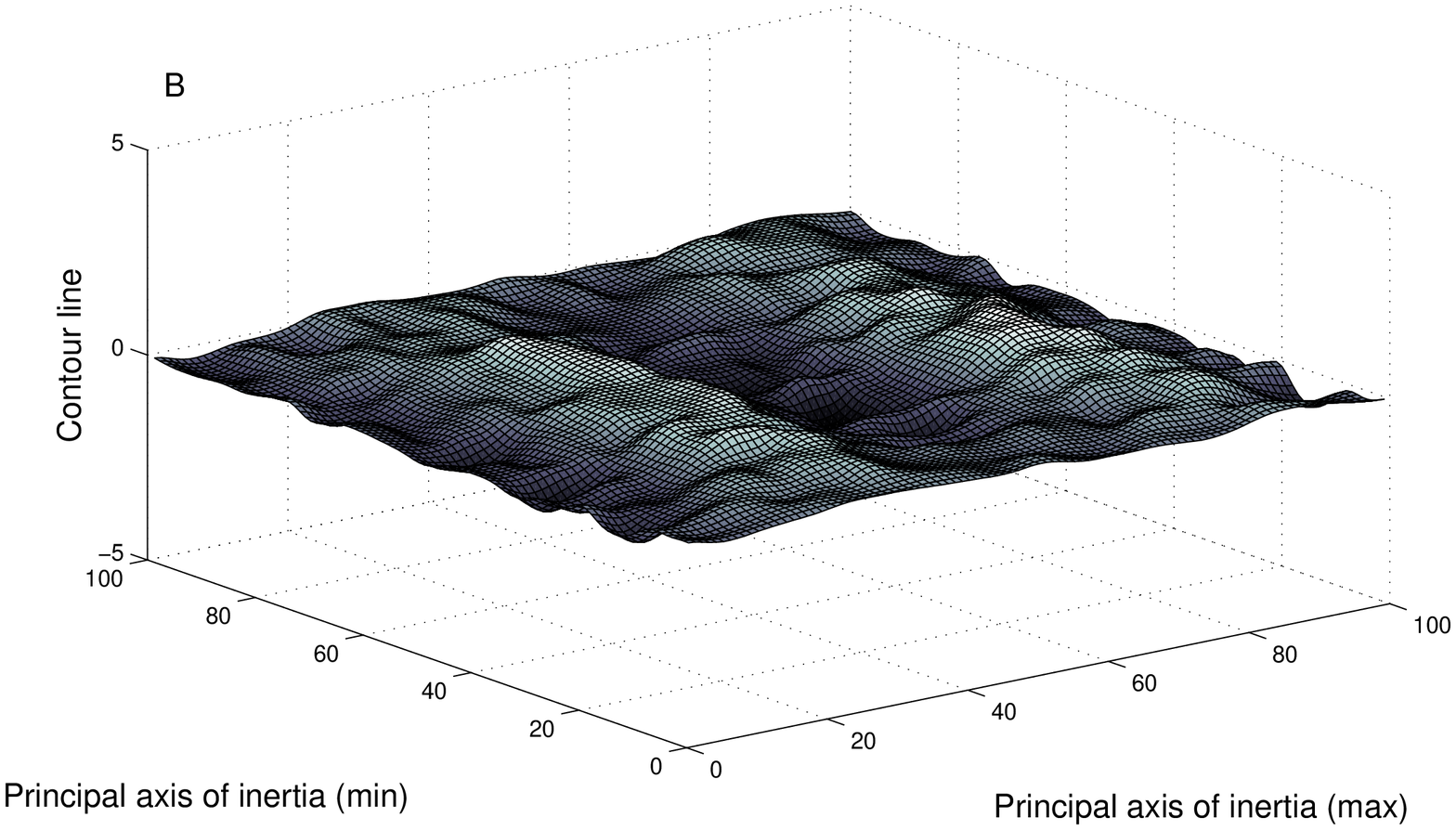}\\
& \includegraphics[width=6.8cm]{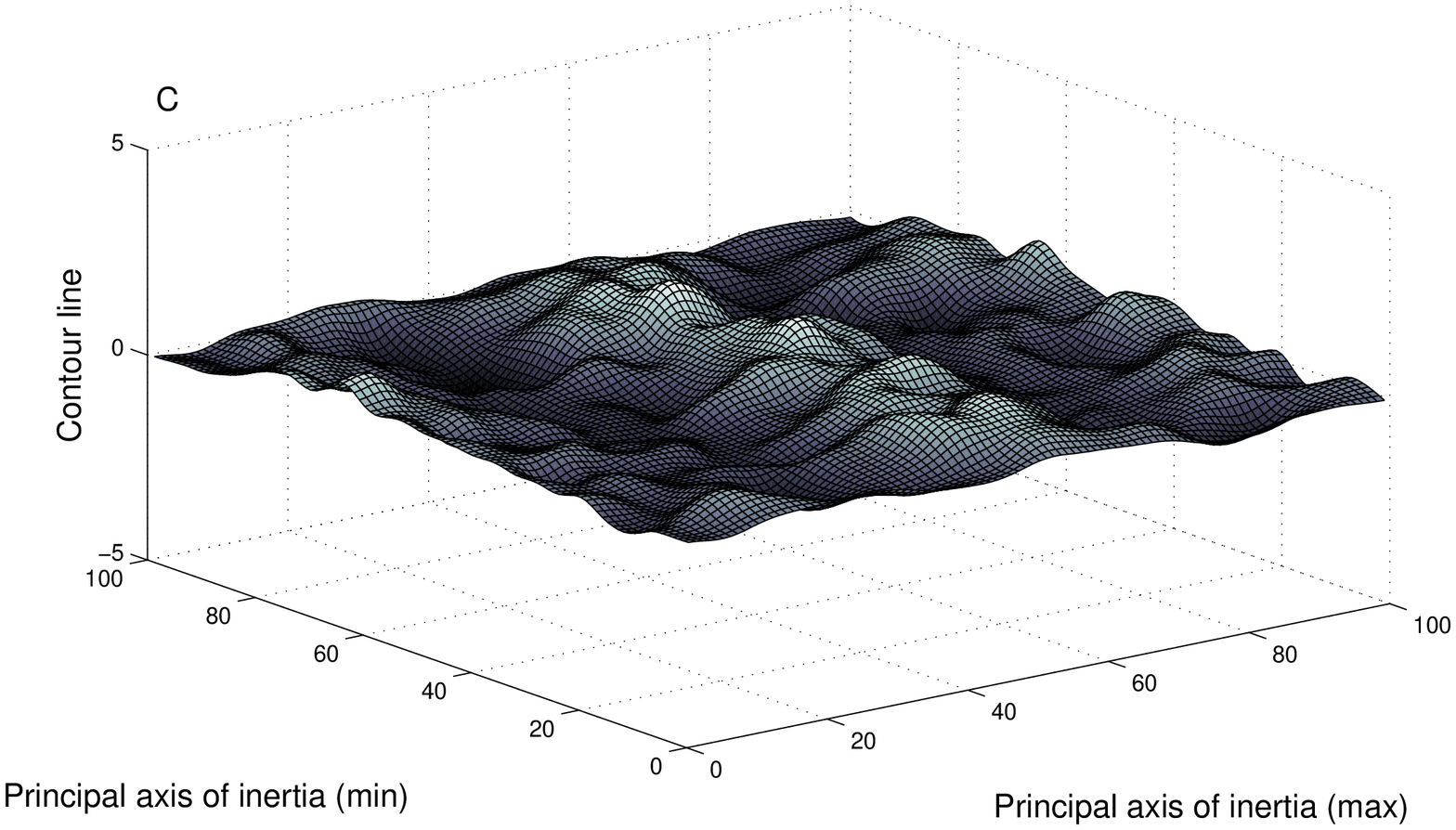}
\end{tabular}
\caption{\label{fig2} Surface mapping of bone. Fig.~\ref{fig2}A Surface mapping of the first measurement of the wrestler right foot's external sesamoid bone beneath the head of the first metatarsal bone. Fig.~\ref{fig2}B Surface mapping of the second measurement of the wrestler right foot's external sesamoid bone beneath the head of the first metatarsal bone. Fig.~\ref{fig2}C Variations of the surface mapping of the wrestler right foot's external sesamoid bone beneath the head of the first metatarsal bone after 18 months (the second measurement). Figs.~\ref{fig2}A, \ref{fig2}B and \ref{fig2}C have been rated by percentage and smoothed. A closed bone surface means that when it is continuous, there is no boundary. But when the surface is cut by the prime meridian, a boundary emerges. So when smoothing the surface, cloud computing method is adopted~\cite{Gu} to ensure the integrity of the object's shape.}
\end{center}
\end{figure}

 The principal axes of sesamoid bone is set up by Eqs.~\ref{eq2} and~\ref{eq3}; the tomography of sesamoid bone is reconstructed after it is rotated by Eq.~\ref{eq4}; the boundary of the sesamoid bone is drawn by Eq.~\ref{eq5}. When the prime meridian is determined by the principal axes, the sesamoid bone surface is developed by Eq.~\ref{eq6} and the mapping of the sesamoid bone surface is accomplished by Eq.~\ref{eq7}. See Fig.~\ref{fig2}.

\begin{figure}[!ht]
\begin{center}
\begin{tabular}{cccc}
 \includegraphics[width=6.8cm]{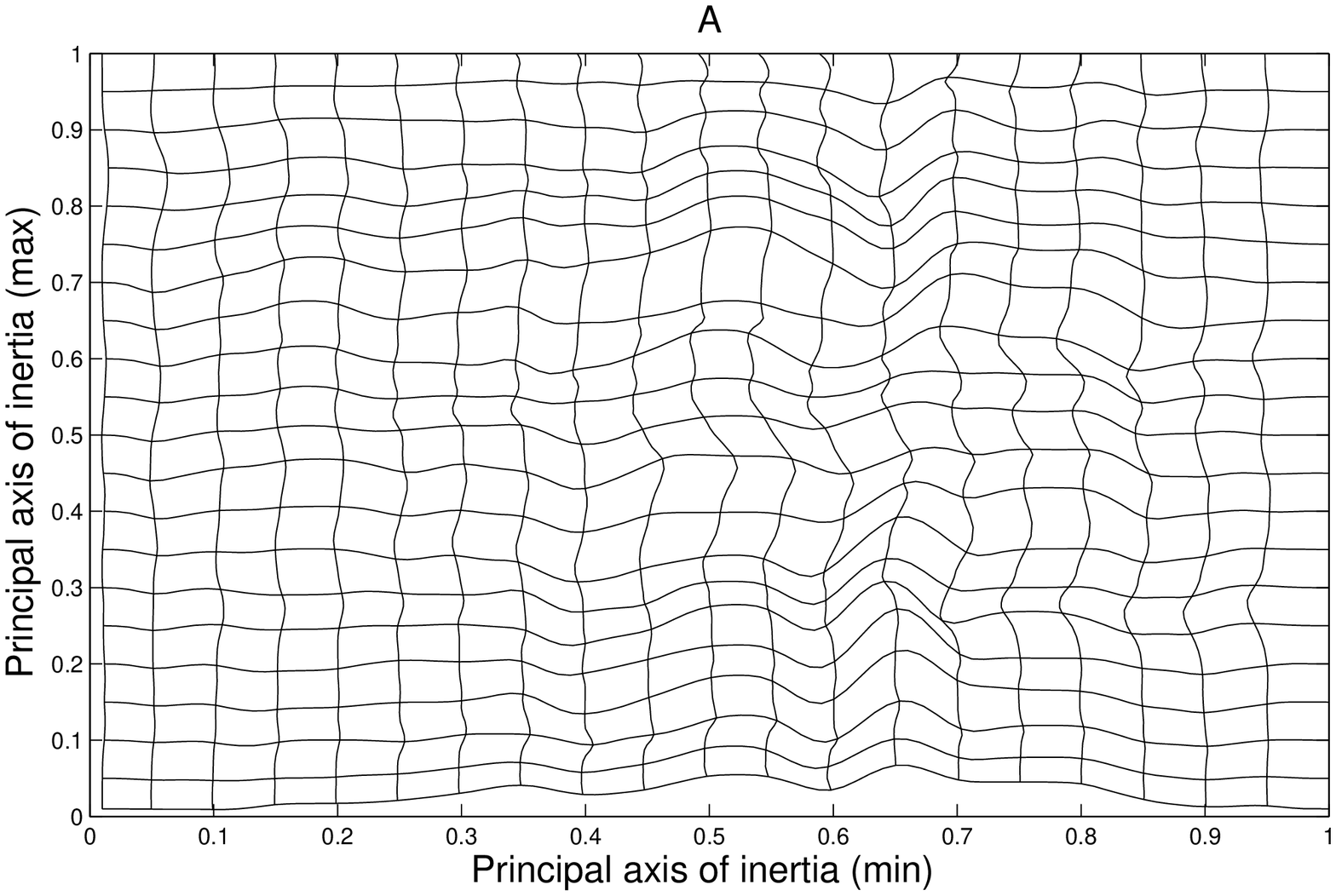}&
 \includegraphics[width=6.8cm]{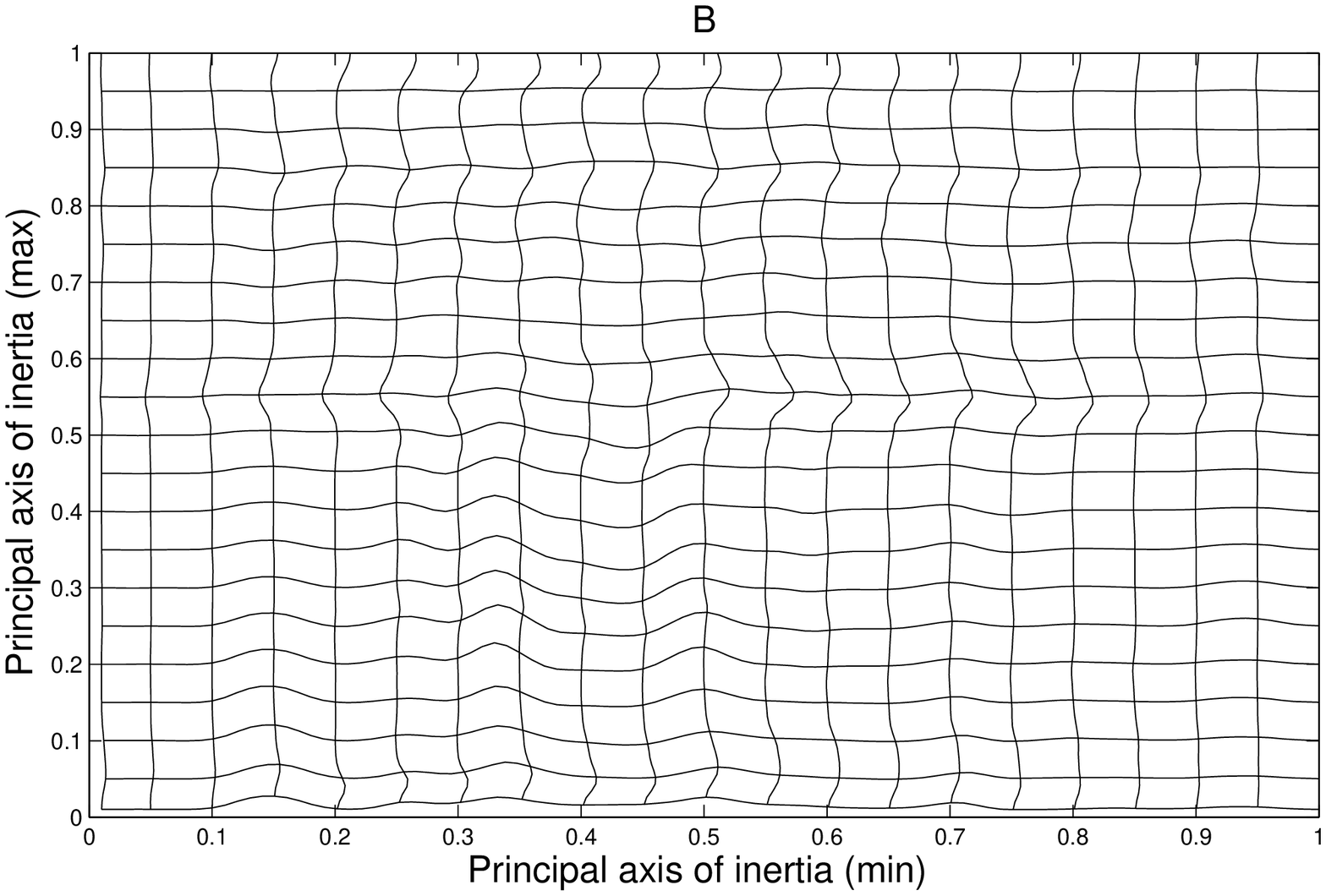}
\end{tabular}
\caption{\label{fig3} Flattened plane of bone.
Fig.~\ref{fig3}A The flattened bone surface of the first measurement of the wrestler's lower external sesamoid bone of the first metatarsal bone.
Fig.~\ref{fig3}B The flattened bone surface of the second measurement of the wrestler's lower external sesamoid bone of the first metatarsal bone after 18 months.}
\end{center}
\end{figure}

Figs.~\ref{fig2}A and 2B indicate that the mapping technique can better illustrate the bone surface properties. Fig.~\ref{fig2}C suggests that the mapping technique can be used as a quantitative method to study the changes of an object's shape.
Using the bone surface mapping, Eq.~\ref{eq8} can flatten the bone surface as a plane. See Fig.~\ref{fig3}.

It can be concluded that the professional training has caused adaptative changes of the sesamoid bone's shape. The structural changes can be depicted by the density and distribution while the shape and its changes can be analyzed by the bone surface mapping.

\section*{Conclusion}
The generalized Papoulis theorem~\cite{Papoulis} elucidates that the bone surface shape keeps its geometric invariance, such as rotation, translation or dimension change~\cite{Nguyen,Rochefort}. This ensures the consistency of the CT scanning results of different postures of bone in vivo when its isotropy is ascertained. The uniqueness of the relative consistency of the inertia reference system of principal moments of inertia on direction cosines provide evidence to the bone surface mapping technique. The characters such as the geometric invariance and the uniqueness of the coordinate system of the principal moments of inertia have enabled the bone surface mapping technique to depict the bone's external morphological characters. This can advance the research of the morphological mechanisms. The experiment of the bone in vivo signifies that the bone mapping technique adds another research method and supplements the analytical method of the bone's three-dimensional imaging technique.

We can understand the world through a map. When the bone surface mapping technique reveals its surface information through a "map", the activities of our life evolve continuously on this map. We anticipate that this mapping technique will be widely used in related disciplines.

\section*{Acknowledgments}

The authors would like to acknowledge the support from the subject.

\newpage

\begin{table}
\caption{\label{t1}Sesamoid bone volume, surface area and density of two measurements}
\small
\begin{center}
\item[]\begin{tabular}{@{}*{7}{l}}
\hline
&\textbf{First}& & &\textbf{Second}& &\\
&volume&surface&density&volume&surface&density\\
&$mm^{3}$&$mm^{2}$&$mg/mm^{3}$&$mm^{3}$&$mm^{2}$&$mg/mm^{3}$\\
\hline
Left - external&$307.06$&$247.41$&$1.84$&$307.87$&$247.96$&$1.85$\\
Left - internal&$204.35$&$184.18$&$1.94$&$200.38$&$183.43$&$1.97$\\
Right - external&$257.84$&$209.59$&$1.87$&$28.83$&$210.73$&$1.92$\\
Right - internal&$205.87$&$181.61$&$1.89$&$203.13$&$181.20$&$1.97$\\

\hline
\end{tabular}
\end{center}
\end{table}

\end{document}